\documentclass[reprint,amsmath,amssymb,aps]{revtex4-2}
\usepackage{graphicx}

\usepackage{dcolumn}
\usepackage{bm}

\usepackage[dvipsnames]{xcolor}

\usepackage{amssymb}
\usepackage{amsmath}
\usepackage{epsfig}
\usepackage{chemarr}
\usepackage{url}
\usepackage{natbib}
\usepackage{color}

\usepackage[]{changes}

\begin{document}

\title{Physical learning of power-efficient solutions}

\author{Menachem Stern$^1$, Sam Dillavou$^1$, Dinesh Jayaraman$^2$, Douglas J. Durian$^{1,3}$ and Andrea J. Liu$^{1,3}$}

\affiliation{$^1$Department of Physics and Astronomy, University of Pennsylvania, Philadelphia, PA 19104}
\affiliation{$^2$Department of Computer and Information Science, University of Pennsylvania, Philadelphia, PA 19104}
\affiliation{$^3$Center for Computational Biology, Flatiron Institute, Simons Foundation, New York, NY 10010, USA}




\date{\today}

\begin{abstract}
As the size and ubiquity of artificial intelligence and computational machine learning (ML) models grow, their energy consumption for training and use is rapidly becoming economically and environmentally unsustainable. Neuromorphic computing, or the implementation of ML in hardware, has the potential to reduce this cost. In particular, recent laboratory prototypes of self-learning electronic circuits, examples of ``physical learning machines," open the door to analog hardware that directly employs physics to learn desired functions from examples. In this work, we show that this hardware platform allows for even further reduction of energy consumption by using good initial conditions as well as a new learning algorithm. Using analytical calculations, simulation and experiment, we show that a trade-off emerges when learning dynamics attempt to minimize both the error and the power consumption of the solution--greater power reductions can be achieved at the cost of decreasing solution accuracy. Finally, we demonstrate a practical procedure to weigh the relative importance of error and power minimization, improving power efficiency given a specific tolerance to error. 
\end{abstract}

\pacs{Valid PACS appear here}
\maketitle

\section{Introduction}

There has been a meteoric rise in the adoption and usage of artificial intelligence (AI) and machine learning (ML) tools in just the last 15 years~\cite{lecun2015deep, shinde2018review}, accompanied by an equally spectacular rise in the sizes of ML models and the amount of computation required to train and apply them~\cite{sze2017hardware, garcia2019estimation}. In recent years, the energy required to train state-of-the-art ML models, as well as to use the trained models, has been rising exponentially, doubling every $4-6$ months~\cite{amodei2018ai}. This energy cost will eventually severely constrain further increases in model complexity and already constitutes a significant economic and carbon cost~\cite{van2021sustainable, gupta2021chasing, wu2022sustainable}.

The field of neuromorphic computing~\cite{mead1990neuromorphic,burr2017neuromorphic, markovic2020physics, schuman2022opportunities} strives to recreate the structure and/or function of the brain in synthetic hardware, in particular the ability to learn in a fashion similar to ML algorithms. A major motivation for the development of neuromorphic systems is the possibility of massive energy savings compared to ML implemented on standard computers~\cite{hasler2013finding}. Many proposals for synthetic `neurons' and `synapses' have been laid out over the past three decades, promising lower power consumption compared to standard computers by $2-5$ orders of magnitude~\cite{sharad2012proposal, schuller2015neuromorphic, kang2020energy, davies2021advancing}. While much neuromorphic computing research has focused on the development of power-efficient hardware, usually for performing inference (applying already-trained ML models), some attention has recently been given to the study of power-efficient learning `algorithms'~\cite{neftci2018data, sorbaro2020optimizing, chakraborty2020pathways, narduzzi2022optimizing}. However, most neuromorphic hardware implementations considered thus far specifically attempt to mimic standard ML algorithms such as backpropagation~\cite{hu_dotproduct_2016, wang_reinforcement_2019, zhang_neuroinspired_2020} or phenomenological neural synaptic learning processes such as STDP (spike-timing-dependent plasticity)~\cite{arima_refreshable_1992,schneider_analog_1993, kim_experimental_2015, labarbera_interplay_2016, serb_unsupervised_2016}. 

Recently, a new avenue was opened toward realizing power-efficient neuromorphic computing, dubbed \textit{physical learning machines} or \textit{self-learning physical networks}~\cite{stern2023learning}. Rather than mimicking known learning algorithms such as backpropagation, such systems exploit their inherent physics in order to learn, using \emph{local learning rules} that modify \emph{learning degrees of freedom} based on locally available information, such that the system globally learns to perform desired tasks. A certain class of local learning rules, known as \emph{contrastive learning}~\cite{movellan1991contrastive, scellier2017equilibrium, stern2021supervised, anisetti2023learning,anisetti2022frequency}, describe how learning degrees of freedom should be modified in order for systems to achieve desired outputs in response to inputs supplied by observed examples of use (\textit{i.e.} supervised learning).

In order to realize any power gains, such learning rules must be implemented in hardware. \emph{Coupled Learning}, a particular contrastive learning rule, has been realized successfully in laboratory hardware for electronic circuits of variable resistors~\cite{dillavou2022demonstration, wycoff2022learning, stern2022physical, dillavou2023circuits}. Such systems already consume less power than conventional computers doing inference because they are analog rather than digital. Here we use analytical theory, computation and experiment to show that the propensity of self-learning electronic circuits to minimize power dissipation enables even greater reductions of power consumption via appropriate initialization and power-efficient learning rules.  We specifically demonstrate these results for regression tasks. It should be noted, however, that our analysis and results should apply to other physical learning machines in different physical media (\textit{e.g.} mechanical networks) if they can be developed in the lab, as well as to other types of problems (\textit{e.g.} classification).

The paper is organized as follows: In Sec. 2 we describe the physical learning approach and discuss how the power consumption of the system is modified by learning, in particular as we change the initial conditions of the learning degrees of freedom. A judicious choice of initial conductances yields learning solutions with low power consumption, while also reducing the energy consumed in training. In Sec. 3 we introduce a modification to the local physical learning rule in order to minimize both error and power consumption. We analyze this new local rule theoretically and test it in simulations and lab experiments, concluding that it leads to an error-power trade-off; lower-power solutions may be obtained at the expense of higher error. The energy required to train the system can be reduced as well. Finally, in Sec. 4 we demonstrate how a power-efficient learning algorithm with dynamical control over the weighting of power and error optimization can lead to efficient adaptation of low-power solutions beyond simply using good initial conditions and constant weighting.


\section{Power consumption in physical learning machines}

In previous work, we established theoretically and experimentally that self-learning resistor networks can be  trained to perform tasks like allostery, regression and classification~\cite{stern2021physical,dillavou2022demonstration,dillavou2023circuits}.  Training a deep neural network corresponds to minimizing a learning cost function with respect to learning degrees of freedom (edge weights and biases). The learning landscape, described by the learning cost function as one axis in the high-dimensional space where each other axis corresponds to a different learning degree of freedom, remains fixed during the minimization. Successful training of physical learning machines, on the other hand, corresponds to simultaneous minimization of \emph{two} cost functions, the learning and physical cost functions, with respect to two different sets of degrees of freedom (DOF), the learning and physical degrees of freedom, respectively. In the case of a self-learning electrical network of variable resistors, the physical cost function is the dissipated power, the physical DOF are the node voltages, and the learning DOF are the conductances. 

Notably, the learning cost function depends implicitly on the physical DOF while the physical cost function depends implicitly on the learning DOF. As a result, both the learning landscape and the physical landscape \emph{evolve} during training. For example, training gives rise to soft modes in the physical landscape as well as stiff modes in the learning landscape, making the system more conductive and lowering its effective response dimension~\cite{stern2023physical}. 

The height of a minimum in the physical landscape corresponds to the power required to actuate the desired response (to obtain the desired outputs in response to the given inputs from training data). Due to the coupling between the learning and physical landscapes, it is possible to find and push down minima in the physical landscape corresponding to global minima in the learning landscape during training, thus decreasing the amount of power required to perform a given task. 

Consider a system that minimizes a scalar physical cost function $P(V;k)$ (\textit{e.g.} the dissipated power) depending on a set of physical DOF $V$ (\textit{e.g.} the node voltages) and a set of learning DOF $k$ (\textit{e.g.} the edge conductances). When an input signal (\textit{e.g.} set of voltages at input nodes) is applied, the system responds by optimizing the physical DOF to minimize $P$, subject to the input constraints, producing a stable free state $V^F$ with an associated physical cost $P^F(V^F;k)$. Training this system for specific output responses using coupled learning~\cite{stern2021supervised} involves clamping the targets $T$ by slightly nudging them toward the desired response $V^C_T=V^F_T+\eta (\tilde{V}_T-V^F_T)$, with $\tilde{V}_T$ the desired response and nudge amplitude $\eta\ll 1$. The physical system then minimizes the physical cost function subject to both the inputs and this clamping, yielding a clamped state $V^C$ with a clamped physical cost $P^C(V^C;k)$. The contrast (or contrastive function) is defined as the difference between the physical cost for the clamped and free states
\begin{equation}
\begin{aligned}
\mathcal{C}\equiv \eta^{-1}[P^C-P^F]
\end{aligned}
 \label{eq:CF},
\end{equation}
which is intrinsically non-negative. Minima with vanishing contrast are also minima of the error (loss) function $\mathcal{L}$ that is typically used to measure the quality of a learning solution, \textit{e.g.} the mean squared difference between the desired and obtained behavior $\mathcal{L}\equiv \frac{1}{2}(\tilde{V}_T-V_T^F)^2$ ~\cite{stern2021supervised}. 

Physical learning is achieved by a learning rule that corresponds to modifying the learning degrees of freedom according to the partial derivative of the contrast. This learning rule is local:
\begin{equation}
\begin{aligned}
\dot{k}= -\alpha \partial_k\mathcal{C}=-\alpha \eta^{-1}\partial_k[P^C-P^F]
\end{aligned}
 \label{eq:LR1},
\end{equation}
with $\alpha$, a scalar \emph{learning rate}, setting the time scale for the learning dynamics. A system following these dynamics with a sufficiently low learning rate tends to minimize a learning cost function $\mathcal{L}$ (see Fig.~\ref{fig:Fig1}b). See Appendix A for more details on the learning dynamics close to a solution for the learning degrees of freedom $k^*$.

\subsection{Power consumption in learned solutions}

We now turn our attention to the scalar physical cost (\textit{i.e.} power consumption) of the free state $P^F(k)$. We first study how the free power is affected by the basic coupled learning rule of Eq.~\ref{eq:LR1}, and we will later see how the free power can be substantially reduced by modifying this rule. Using the chain rule on \ref{eq:LR1}, we can derive an ODE for the free state power during training
\begin{equation}
\begin{aligned}
\dot{P}^F(k)=\dot{k}^T \nabla_k P^F(k) = -\alpha \partial_k\mathcal{C}^T\cdot \partial_k P^F
\end{aligned}
\label{eq:EF0}.
\end{equation}
Note that the free state is a fixed point of the physical dynamics, so that the derivative $\partial_V P^F$ vanishes exactly, and hence $\nabla_k P^F=\partial_k P^F+\frac{dv}{dk}\partial_V P^F=\partial_k P^F$. We see that the free state power tends to decrease if the gradients of the power and the contrast w.r.t $k$ align, and increase otherwise. Assuming the power changes slowly with $k$, or that the learning DOF $k$ are close to the learning solution $k^*$, we can approximate the free state power using Taylor expansion 
\begin{equation}
P^F(k)\approx P^F(k^*)+(k-k^*)^T \partial_k P^F(k^*).
\end{equation}
This shows that the free power consumption changes due to the learning dynamics, starting at the initial condition $P^F(k^0)$ and ending after training with $P^F(k^*)$. We next discuss the sign of this power shift, determined by the alignment between the gradients of the contrast and free state power. 

Let us specialize to the case of linearized resistor networks, where the physical DOF are the voltages at nodes $V_a$, while the learning DOF are conductances $k_{i}$ of edges $i$ connecting pairs of nodes. An adjacency matrix $\Delta_{ia}$ is defined such that each row of the matrix corresponds to an edge, having a value of $+1$ at index of the incoming node of that edge, $-1$ at the index of the outgoing node, and $0$ elsewhere. The choice of which node is incoming or outgoing is a matter of convention and sets the direction of currents but has no physical consequence. The vector of voltage drops on edges is given by $\Delta V_i = \sum_a \Delta_{ia} V_a$. Resistor networks minimize the total power dissipation
\begin{equation}
\begin{aligned}
P &= \frac{1}{2} \sum_{i} k_i \Delta V_i^2 = \frac{1}{2}\sum_{abij} V^T_a \Delta^T_{ai} K_{ij} \Delta_{jb} V_b
\end{aligned}
 \label{eq:ResNetPower},
\end{equation}
with $K_{ij}$ a diagonal matrix whose diagonal elements are $k_i$. In such networks, where one of the nodes is grounded at $V_G=0$, the native state of the network (in the absence of any inputs) is where all voltage values are zero, all voltage drops are zero and the total power dissipation is $P=0$. 
When the free\textbackslash clamped boundary conditions are applied, \textit{e.g.} by introducing currents in certain input and output edges, the free and clamped power are

\begin{equation}
\begin{aligned} \nonumber
P^{F,C} &=  \frac{1}{2}\sum_{abij} (V^{F,C})^T_a \Delta^T_{ai} K_{ij} \Delta_{jb} (V^{F,C})_b
\end{aligned}
 \label{eq:Energies}.
\end{equation}

Given weak clamping ($V^C-V^F\sim\eta\ll 1$), we can write the contrast function $\mathcal{C}$, neglecting terms proportional to $\eta^2$:
\begin{equation}
\begin{aligned}
 \mathcal{C} \approx & \frac{1}{2} \alpha \eta^{-1}\sum_{ab}\{(V^C-V^F)_a V^F_b +\\
&+ V^F_a (V^C-V^F)_b \}\sum_{ij} \Delta^T_{ai} K_{ij} \Delta_{jb} ,
\end{aligned}
 \label{eq:Contrast}
\end{equation}
In the curly brackets, we see the appearance of a rank-1 symmetric matrix, formed from an outer product of the free response and clamped response. We take the partial derivative of the contrast w.r.t $k$~\cite{anisetti2023learning}:
\begin{equation}
\begin{aligned}
\frac{\partial \mathcal{C}}{\partial k_{i}}  \approx  \alpha \eta^{-1}  [\Delta (V^C-V^F)]_i [\Delta V^F]_i
\end{aligned}
 \label{eq:ContrastGrad}.
\end{equation}



In this simple case, the learning modification is determined by the alignment of each component of the free state response $\Delta V^F_i$ with its nudge in the clamped state $(\Delta V^C - \Delta V^F)_i$. In these particular models, we also know that the the free state power gradient is positive $\frac{\partial P^F}{\partial k_{i}} = (\Delta V^F)_i^2 \geq 0$. We conclude that if the clamped state nudge aligns with the free state response, the free state power will tend to decrease. This is sensible as the system has to decrease its conductances to achieve a stronger response required by the clamping. The opposite effect occurs when the nudged response is misaligned with the free state, resulting in increased conductances.  

\subsection{Power dependence on initial conditions}

In Sec. IIA we established how physical learning affects the system's free state power consumption. In the following we consider how the initial conditions of the learning degrees of freedom determine the free state power of the learned solutions.

It is well recognized in the ML literature that the dynamics and obtained solutions of learning algorithms strongly depend on initialization, \textit{i.e.} the initial values of the learning DOF~\cite{fernandez2001weight,sutskever2013importance,bahri2020statistical,narkhede2022review}. In the context of physical learning, the choice of initialization may not only affect the training time and accuracy of a solution but may also have important effects on the power required to actuate the system in the obtained learning solution. Suppose a set of voltage drops is applied over some input edges of a resistor network, and we read out the resulting voltage drops over some other output edges. Also suppose that the conductance values of the network have a certain scale $\kappa$. It is known that the output voltage drops do not depend on the scale $\kappa$, but only on the relative ratios of the conductance of different edges. However, reducing the conductance scale does in fact linearly decrease the dissipated power associated with the free state (Eq.~\ref{eq:ResNetPower}). We can thus in principle improve the power dissipation indefinitely by reducing the conductance scale. Realistically, we are bound by experimental considerations: variable conductive elements have minimal conductance values (corresponding to maximal resistance). Furthermore, low conductance necessitates more precise hardware implementations, as the network response becomes highly sensitive to small variations in the conductance.

\begin{figure}
\includegraphics[width=0.95\linewidth]{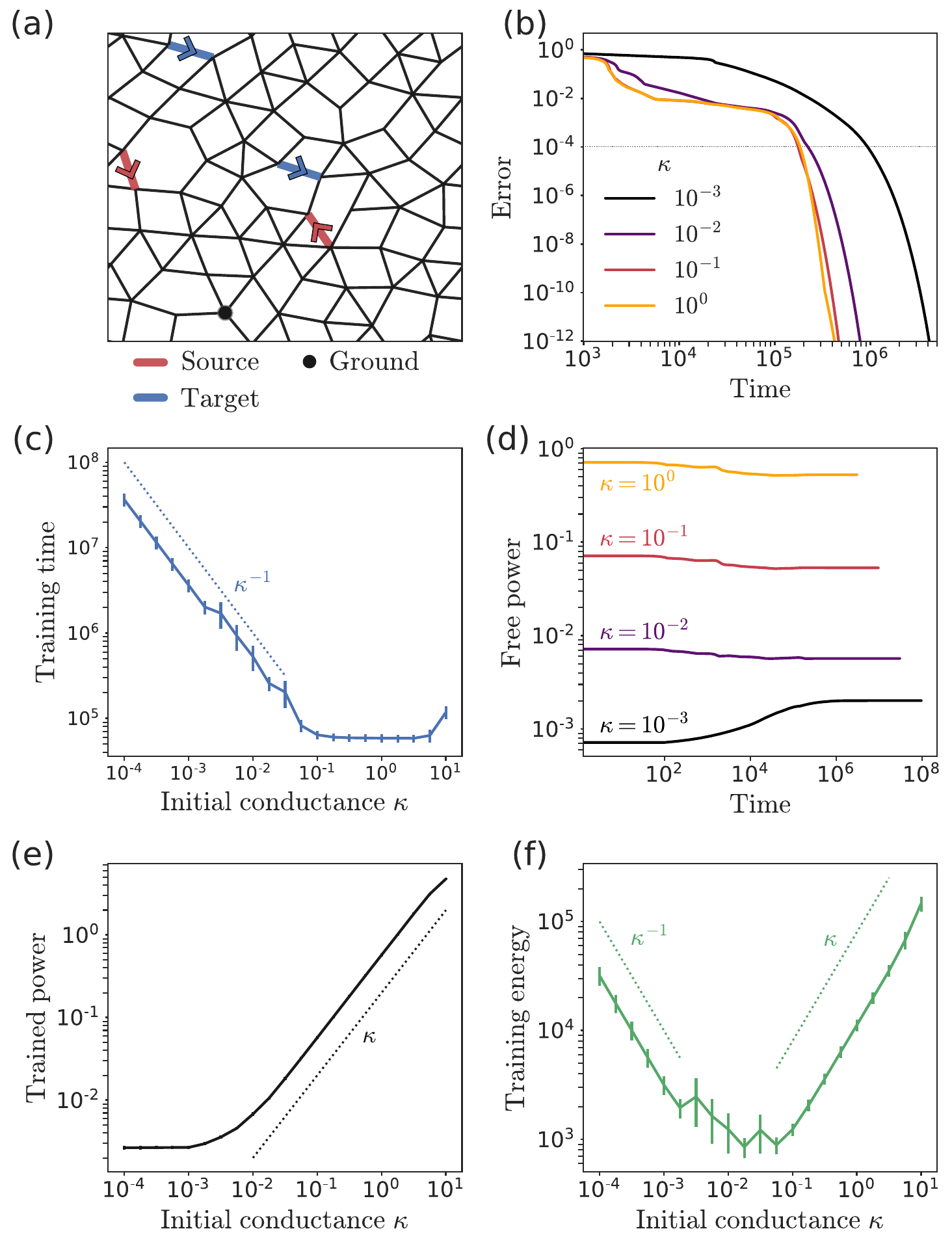}
\caption{Effects of varying the conductance initialization scale. a) Simulated resistor networks with edges corresponding to variable resistors. We train networks with $N=64$ nodes to perform linear regression, \textit{i.e.} to simulate desired linear equations with $2$ variables (red source edges) and $2$ results (blue target edges). b) Coupled learning successfully trains these networks, reducing the error $\mathcal{L}$ by multiple orders of magnitude. Changing the conductance initialization scale $\kappa$ has little effect on the success of training. c) The training time (time taken for the error to drop to a certain level $\mathcal{L}=10^{-4}$ remains constant when initialization is far from the bounds, but grows linearly for low initialization close to $k_{\rm{min}}$. d,e) Decreasing the conductance initialization scale has a strong effect, reducing the power required to actuate the learned solution. f) Choosing a proper optimal initialization, we can reduce both the power of the obtained solution and energy required to train the network. Results averaged over 50 realizations of networks and regression tasks.
\label{fig:Fig1}}
\end{figure}

The above considerations suggest that initializing the conductance values $k$ (learning DOF) at lower values may yield solutions with lower dissipated power. 
To verify these ideas, we trained $N=64$ node networks (Fig.~\ref{fig:Fig1}a) for multiple regression tasks with $2$ inputs and $2$ outputs (see appendix B for details on the simulated resistor networks and regression tasks). We initialized the conductance values uniformly with different conductance scales in the range $10^{-4}\leq \kappa\leq 10^1$. We note that in these simulations, the minimum conductance for any given edge is $k_{\rm{min}}=10^{-4}$, and learning modifications that attempt to lower the conductance below $k_{min}$ are not performed. 
The learning rate $\alpha$ has been chosen such that at $\kappa=1$, the learning rate is $\alpha=0.33$, a value which typically results in relatively quick and stable learning performance for these networks and tasks. The networks are trained for $10^6$ learning iterations of Eq.~\ref{eq:LR1}. As expected, we find that coupled learning reduces the error by many orders of magnitude (Fig.~\ref{fig:Fig1}b). We also find that when the learning rate is scaled appropriately $\alpha\propto \kappa$, the training time (time taken for the system to reach a certain error threshold $\mathcal{L}=10^{-4}$) does not change much at relatively high initialization $\kappa$ (Fig.~\ref{fig:Fig1}c). However, initialization close to the lower boundary $k_{\rm{min}}$ induces a linear increase in the training time, scaling as $\kappa^{-1}$. This increase in training time is reasonable as a large part of the training modification $\Delta k_i$ is not performed because it would require the conductances to go below the minimum. As a result, the learning rate is effectively lowered. Note that in all simulations, the unit of time is defined as an epoch, i.e. the time taken for the network to observe all of the training examples associated with a task and modify its learning degrees of freedom according to the coupled learning rule.


More importantly, at lower initialization scales, physical learning finds lower power solutions (Fig.~\ref{fig:Fig1}d, e). These results clearly show the benefit of initializing the conductances of edges close to their minimal values in terms of learning power efficient solutions. While we so far referred to the power necessary to actuate the solution (\textit{i.e.} the free state power), one often needs to consider the energy required to train the network to adopt this solution. In some applications, the energy required to train a system is small compared to the total energy spent to use it throughout its life cycle. However, when this is not the case, one should consider learning algorithms that reduce the required training energy as well as the free power. In our simulations, the energy required to train the system can be measured as the integral over the free state power during training, until the error reaches a certain tolerable level (\textit{e.g.} $\mathcal{L}=10^{-4}$). We find that the training energy scales linearly with the initialization at high $\kappa$ (Fig.~\ref{fig:Fig1}f),  similar to the free power of the obtained solution. However, lowering $\kappa$ close to $k_{\rm{min}}$ actually \emph{increases} the training energy. This is because we can no longer realize gains in the free power (plateau region in Fig.~\ref{fig:Fig1}e) while the training time increases linearly with decreasing $\kappa$ (Fig.~\ref{fig:Fig1}c). As a result, the training energy in this regime increases linearly with decreasing $\kappa$ (Fig.~\ref{fig:Fig1}f). Thus there is an optimal value for the initialization $\kappa$ corresponding to the minimum training energy.

In machine learning, however, the greatest energy cost is incurred during inference. In our case, this cost is quantified by the trained power. We note that training reduces the free power for high $\kappa$, but increases it for low $\kappa$ next to the lower conductance limit (Fig.~\ref{fig:Fig1}d). This is sensible, because for low initial conductances at or near the minimum, the network must increase some edge conductances in order to decrease its error. That said, we conclude that initializing the network with properly low conductance values can save significant energy during learning and when using the trained network.

\section{Explicit power minimization}

We have seen that the learning rule of Eq.~\ref{eq:LR1} modifies the power of the free state during learning. Our next step is to find a way to explicitly control the power consumption when inputs are applied. This is possible because the learning rule of Eq.~\ref{eq:LR1} is already written in terms of the power consumed by the system. It is natural to modify this learning rule to locally minimize this power as well as the error. Consider the addition of an explicit power minimization term to the contrast:
\begin{equation}
\begin{aligned}
\mathcal{C}_\lambda= \eta^{-1}[P^C-P^F]+ \eta^{-1}\lambda P^F ,
\end{aligned}
 \label{eq:CFL}
\end{equation}
where $\lambda$ is a tunable parameter that dictates the importance of power minimization. The learning rule, the partial derivative of the contrast, then becomes:
\begin{equation}
\begin{aligned}
\dot{k}=-\alpha\partial_k \mathcal{C}_\lambda = -\alpha \eta^{-1}\partial_k[P^C- (1-\lambda) P^F] .
\end{aligned}
 \label{eq:CFLD}
\end{equation}
Note that as the free state power can be partitioned as a sum over the network edges, the power minimizing rule is still local and physically realizable. 
This modified learning rule tends to decrease the free state power, as the modified learning dynamics lower the free power \textit{and} the contrast of Eq.~\ref{eq:CF}. If we set $\lambda =1$, the free state power cancels out and we recover the directed aging learning rule~\cite{pashine2019directed,hexner2019effect} that solely tends to reduce the power of the clamped state. 

Using the modified learning rule (Eq.~\ref{eq:CFLD}), one can derive ODEs for the contrast and free state power, similar to Eq.~\ref{eq:EF0}:

\begin{equation}
\begin{aligned}
\dot{\mathcal{C}}(k)& = -|\partial_k \mathcal{C}|^2 - \lambda \partial_k\mathcal{C}^T \partial_k P^F \\
\dot{P^F}(k)& = -\partial_k\mathcal{C}^T \partial_k P^F - \lambda |\partial_k P^F|^2
\end{aligned}
 \label{eq:RegODEs}
\end{equation}

These dynamics tend to reduce the value of the contrast $\mathcal{C}$ over time, up to interference from a term that encodes the alignment between the gradient of the contrast and free state power. Moreover, we find that the free state power tends to be reduced by these dynamics, again up to an effect determined by the alignment. We now discuss the dynamics of the contrast and free state power in a simplified linear setting. First, note that in the limit $\lambda\rightarrow\infty$, the learning rule minimizes the free state power. We denote this free state power minimum as $k^*_\infty$. Around this local minimum, the free state power can be expanded to quadratic order, $$P^F(k)\approx P^F(k^*_\infty)+\frac{1}{2}(k-k^*_\infty)^T H (k-k^*_\infty),$$ where $H\equiv\partial^2_k P^F(k^*_\infty)$ is the free state power Hessian w.r.t to learning degrees of freedom. We can similarly expand the contrast in series around the $\lambda=0$ learning solution (the unmodified solution discussed earlier), $$\mathcal{C}(k)\approx \frac{1}{2}(k-k^*_0)^T \mathcal{H} (k-k^*_0),$$ where $\mathcal{H}\equiv\partial^2_k \mathcal{C}(k^*_0)$ is the contrast Hessian w.r.t to learning degrees of freedom at $\lambda=0$ (in overparameterized networks the constant term $\mathcal{C}(k^*_0)$ vanishes, see appendix A). If the learning solution at finite $\lambda$, $k^*_\lambda$ is close to the limiting solutions $k^*_\infty$ and $k^*_0$, we can express the new contrast approximately as

\begin{equation}
\begin{aligned}
\mathcal{C}_\lambda(k)&\approx \frac{1}{2}(k-k^*_0)^T \mathcal{H} (k-k^*_0) +\\
&+\lambda [P^F(k^*_\infty)+\frac{1}{2}(k-k^*_\infty)^T H (k-k^*_\infty)]
\end{aligned}
 \label{eq:RegCont}
\end{equation}

We can now discuss the dynamics of the learning degrees of freedom $\dot{k}=-\partial_k \mathcal{C}_\lambda$. Taking the partial derivative of Eq.~\ref{eq:RegCont}, we find a first order ODE for $k$, whose solution is exponential:

\begin{equation}
\begin{aligned}
k(t)&=k^*_\lambda  + e^{-(\mathcal{H}+\lambda H)t}[k(t=0) - k^*_\lambda]\\
k^*_\lambda&=(\mathcal{H}+\lambda H)^{-1} [\mathcal{H}k^*_0 + \lambda H k^*_\infty]
\end{aligned}
 \label{eq:WExpSol}
\end{equation}

Starting from an initial condition $k(t=0)\equiv k^0$, the learning DOF exponentially decay to $k^*_\lambda$. Let us discuss the learning DOF solution $k^*_\lambda$. It is clear that when no power optimization is applied $k^*_\lambda=k^*_0$. If both Hessians $\mathcal{H},H$ are full rank (and invertible), the $\lambda$ parameter would smoothly interpolate between $k^*_0$ and $k^*_\infty$. However, we know that the Hessian of the contrast in over-parameterized learning machines is low-rank (with the number of non-zero eigenvalues equal to the number of training tasks, see Appendix A for details)~\cite{stern2023physical}. This means the contrast Hessian $\mathcal{H}$ is not invertible and has vanishing eigenvalues. In the eigen-directions of these vanishing eigenvalues, the power minimization is dominant for any finite value of $\lambda$. Thus, the power minimization term introduces a singular perturbation, so that for infinitesimal power optimization amplitude $\lambda=0^+$ the learning solution approaches $k^*_{0^+} =  \lim_{\lambda\to 0} k^*_\lambda  \ne k^*_0$. The solution $k^*_{0^+}$ tends to minimize the free state power while keeping the contrast low. For over-parameterized learning in the $\lambda \rightarrow 0$ limit

\begin{align}
\begin{aligned}
&k^*_{0^+}= \underset{k}{\text{argmin}}\ P^F(k)\\
& \text{s.t.}\ \  \mathcal{C}(k)=0
\end{aligned}
\label{eq:WEps}
\end{align}

The solution $k^*_\lambda$ is then a weighted average of the limiting solutions $k^*_{0^+},k^*_\infty$, weighted by the Hessian matrices $\mathcal{H}, \lambda H$. For weak power optimization ($\lambda \ll 1$), 

\begin{align}
\begin{aligned}
k^*_\lambda &\approx k^*_{0^+} + \lambda s\\
s&\equiv (\mathcal{H}+\lambda H)^{-1} H(k^*_\infty-k^*_{0^+})
\end{aligned}
\label{eq:other}
\end{align}

For $\lambda\ll 1$, note that the vector $s$ is nearly constant, as the inverse matrix is dominated by $\mathcal{H}$. This means the solutions shift $\lambda s$ is approximately linear in the optimization parameter $\lambda$ (See appendix A). Let us further denote $\Delta k^0=k^0-k^*_\lambda$ and introduce a time propagator $U_\lambda(t)\equiv e^{-(\mathcal{H}+\lambda H)t}$.
The solution for $k$ can be plugged in the equations above to express the dynamics of the contrast and free state power:

\begin{equation}
\begin{aligned}
\mathcal{C}(t)&\approx\frac{1}{2}\Delta k^{0T} U_\lambda\mathcal{H}[U_\lambda\Delta k^0 + 2\lambda s] +
 \frac{1}{2}\lambda^2 s^T\mathcal{H}s \\
P^F(t)&\approx P^F_{0^+} +  \frac{1}{2}\Delta k^{0T} U_\lambda [ H U_\lambda\Delta k^0 + 2\partial_k P^F_{0^+}]  - \\
&-\lambda (\partial_k P^{F}_{0^+})^T (\mathcal{H}+\lambda H)^{-1} [H U_\lambda\Delta k^{0}  + \partial_k P^{F}_{0^+}] 
\end{aligned}
 \label{eq:Solutions}
\end{equation}

For both the contrast and free state power, we keep the largest non-vanishing contribution at long times due to the modified learning dynamics.

\begin{equation}
\begin{aligned}
\mathcal{C}(t\rightarrow\infty)&\approx \frac{1}{2}\lambda^2 s^T\mathcal{H}s \\
P^F(t\rightarrow\infty)& - P^F_{0^+}\approx  \lambda (\partial_k P_{0^+}^{F})^T (\mathcal{H}+\lambda H)^{-1} \partial_k P_{0^+}^{F}
\end{aligned}
 \label{eq:SolutionsInf}
\end{equation}

This is our key result. The error induced by power minimization scales with $\lambda^2$ while the free state power reduction compared to $P^F(k^*_{0^+})$ scales linearly with $\lambda$. 

Our argument considers the error and power of the learned solutions at infinite time, but a practical learning scenario ends after some finite training time $t=\tau$. This training time must be large compared to the natural scale of the contrast Hessian to allow learning to occur. However, in the weak power minimization limit, this time can be much smaller than the power minimization timescale $\mathcal{H} \gg \tau^{-1} \gg \lambda H $. In this case, the dynamics can be approximated by a fast decay towards the unmodified solution $k^*_0$, followed by a slow decay from $k^*_0$ to the power minimizing solution $k^*_\lambda$. For small $\lambda$, the learned solution at a finite time $\tau$ is

\begin{equation}
\begin{aligned}
k(\tau)-k^*_0\approx  \lambda \tau \times H (k^*_\lambda - k^*_0).
\end{aligned}
 \label{eq:SolutionsTau}
\end{equation}

The learned solution moves away from $k^*_0$, at a rate linearly proportional to $\lambda$. This solution can be used to estimate both the contrast and free state power at time $\tau$. As seen before, we find that the contrast scales as $\lambda^2$, while the free state power is reduced proportional to $\lambda$ and the elapsed time $P^F(\tau)-P^F_0\sim - \lambda\tau$.


\begin{figure}
\includegraphics[width=0.95\linewidth]{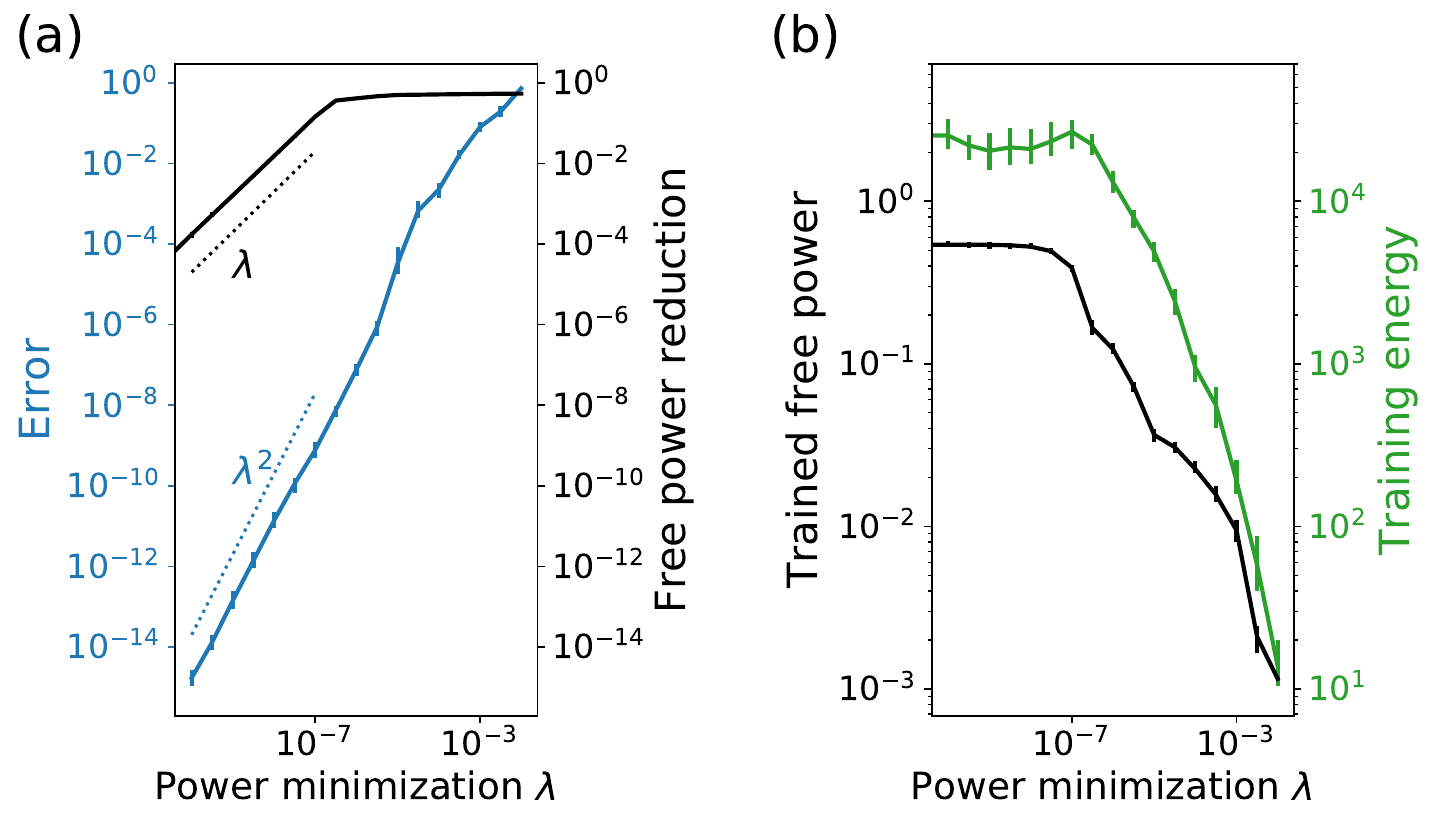}
\caption{Physical learning with power minimization. a) As the relative amplitude of power minimization $\lambda$ is increased, the error of the learned solution increases quadratically, but the power of these solutions is linearly decreased. b) The power of the learned solutions, as well as the training energy necessary to obtain these solutions, decreases with $\lambda$, underscoring a trade-off between power-efficiency and error.
\label{fig:Fig2}}
\end{figure}

Overall, these considerations suggest that under $\lambda$-modified dynamics a trade-off emerges between the error and free state power of the trained solution.
This intuition is verified in numerical simulations in Fig.~\ref{fig:Fig2}. We train a $64$ node resistor network, initialized with intermediate conductance values $k^0_i=1$, for a regression task as before. Here, the training proceeds with the modified power minimization learning rule (Eq.~\ref{eq:CFLD}), varying $\lambda$ in the range $10^{-10}\leq\lambda\leq 10^{-2}$. We train these networks for $\tau=10^5$ steps, and then measure the trained error and free state power, averaging the results over $50$ realizations of the network and regression tasks. We find that for small $\lambda$, the error and free power reduction scale as predicted by Eq.~\ref{eq:SolutionsInf} (Fig.~\ref{fig:Fig2}a). As before, we can compute the total energy required to train these networks. We plot the free power obtained by these trained network and the energy required to train them in Fig,~\ref{fig:Fig2}b. Both of these are markedly decreased when the relative minimization amplitude $\lambda$ is increased, showing the predicted trade-off between power-efficiency and error.

\subsection{Experimental results}

\begin{figure}
\includegraphics[width=0.95\linewidth]{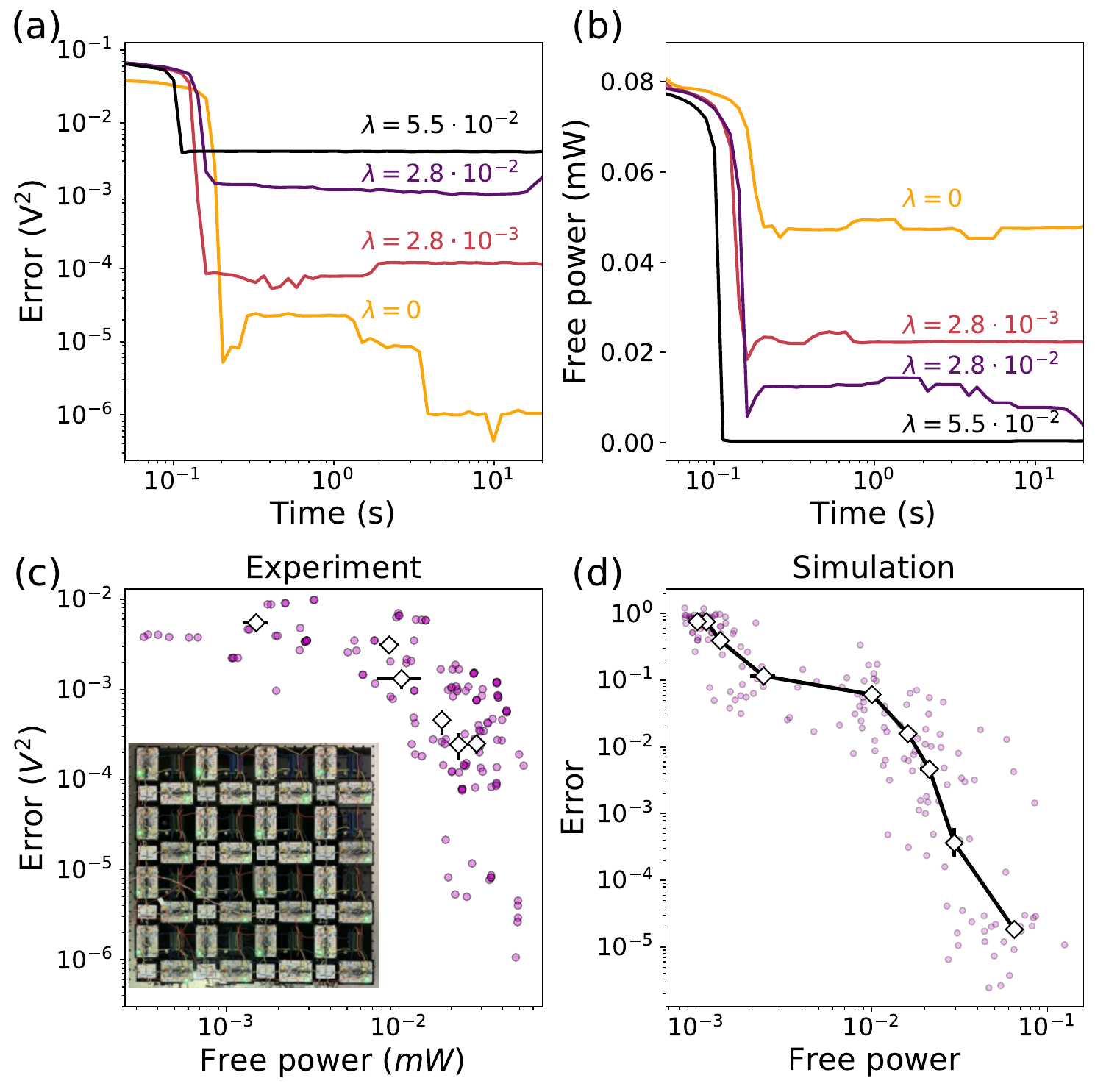}
\caption{Experimental results for power optimization show a trade-off between error and power. a) An experimental network of adaptive nonlinear resistors can physically learn to adopt desired function. This network learns to perform allostery tasks, gradually minimizing the error down to a finite error floor. Error dynamics are shown for different values of the power minimization amplitude $\lambda$.
b) As experiments are run with increasing power minimization ($\lambda$), the learning process finds solutions with increasing error but improved power efficiency. c) Overall, we experimentally observe an error-power trade-off in this experimental learning machine. Inset shows a photograph of the experiment. d) This trade-off between power efficiency and error is recapitulated in simulated learning resistor networks. 
\label{fig:Fig3}}
\end{figure}

So far we argued on theoretical grounds that error can be traded-off for power efficiency by employing the learning rule in Eq.~\ref{eq:CFLD} and verified that in simulations. Here we verify the existence of the trade-off in laboratory experiments. 
We use an experimental network of variable resistors implementing coupled learning, similar to realizations in previous works ~\cite{dillavou2022demonstration, wycoff2022learning, stern2022physical}. However, in this new implementation of the experiment, transistors replace the digital potentiometers in the role of variable resistors~\cite{dillavou2023circuits}. Unlike previous work~\cite{dillavou2022demonstration}, this system is also able to learn according to the continuous coupled learning rule (Eq.~\ref{eq:LR1}), as each resistance element is set by a charged capacitor (on the gate of the transistor) instead of a discrete counter. Modifications to the learning rule of the form of Eq.~\ref{eq:CFLD} are achieved by varying the measurement amplification from the free and clamped networks. Also unlike previous implementations, this new network operates continuously in time, with the clamped state value updated automatically via an electronic feedback loop, and so training duration is measured in real time rather than training steps. Because of unavoidable noise in the experiment, $\eta \rightarrow 0$ is unobtainable; as the clamped state approaches the free state their difference becomes more and more difficult to measure. We therefore use a finite value $\eta = 0.22$ for these experiments, with an effective learning rate of $\alpha = \frac{1}{24 ms}$. Experiments lasted 20 seconds each, and the network's resistances had completely settled at the end of each run. The network is a 4x4 square lattice of edges (inset in Fig.~\ref{fig:Fig3}c) with periodic boundary conditions; edges are initialized with uniform conductance in the approximate middle of their range at the start of each experiment.


The network was trained for 150 two-source, two-target allostery tasks, wherein the sources were held at the low and high end of the allowable range ($0$ and approximately $0.45V$, respectively), with the two desired target outputs at either $20$ and $80\%$ or at $10$ and $90\%$ of this range, respectively. Across these experiments, $\lambda$ was varied to $7$ values ranging $0-0.055$. In all cases the network was able to lower the error, as shown for typical error vs training time curves in Fig.~\ref{fig:Fig3}a. For these tasks, the network also consistently lowered the power of its free state, as shown for the complementary power curves over training time in Fig.~\ref{fig:Fig3}b. Consistent with theoretical predictions, error and power respectively increased and decreased with increasing $\lambda$, with their trade-off shown in Fig.~\ref{fig:Fig3}c. White diamonds correspond to the mean error and free power of all experiments performed with the same value of $\lambda$.

To study this trade-off seen in the experiment, we simulated $N=64$ node resistor networks as done earlier with the addition of a Gaussian white noise term to Eq.~\ref{eq:CFLD} with scale $\delta=5\cdot 10^{-4}$ to approximate the noisy conditions of experimental learning. The white noise term leads to an error floor $\mathcal{L}\sim 10^{-5}$, similar to the experiments. The results for error and free power with $\lambda$ in the range $10^{-6}-10^{-2}$, averaged over $50$ realizations of the network and tasks, are shown in Fig.~\ref{fig:Fig3}d and qualitatively show the same error-free power trade-off.


\section{Dynamical control for greater power minimization}

In the previous section we showed how adding an explicit power minimization term in the contrast function leads to a new local learning rule that attempts to minimize both the error and free state power at the same time, leading to a trade-off between them controlled by the power minimization amplitude parameter $\lambda$. We note that noisy inputs make it impossible to reach zero training error, and in any case, there is experimental noise in the self-learning circuits, so there is a nonzero error floor in practice. Here we use this insight to design a practical control scheme to dynamically modify $\lambda$ during learning, in order to attain tolerable error with more power-efficient solutions . We will show how such a control scheme can yield even more power-efficient solutions compared to using a smart initialization (as in Sec. II) and constant $\lambda$ (as in Sec. III).

Assume we initialize the conductances of a resistor network at their minimal value (maximum resistance). This initialization leads to a free state $V^F(k_{\rm{min}})$ with the lowest possible power dissipation $P^F_{\rm{min}}$. This state corresponds to the minimum power found by the power minimization dynamics with $\lambda \gg 1$, which selects the learning degrees of freedom resulting in the lowest power $P^F_{\rm{min}}$. As seen in Fig.~\ref{fig:Fig2}, reducing the amplitude $\lambda$ from infinity toward zero monotonically decreases the error while increasing the solution free power. 

Here we consider a simple dynamical control scheme. Briefly, we set a specific error tolerance as a target, $\tilde{\mathcal{L}}$. We measure the instantaneous error $\mathcal{L}$ while learning using the local rule Eq.~\ref{eq:CFLD}. If the instantaneous error is larger than the desired tolerance, we decrease $\lambda$ to promote error minimization, while if the error is smaller than the tolerance, we increase $\lambda$ to emphasize power minimization. In other words,

\begin{equation}
\begin{aligned}
\dot{\lambda} = \rho^{-1} \Bigl[\Bigl( \frac{\tilde{\mathcal{L}}}{\mathcal{L}} \Bigr)^p -1 \Bigr] \lambda 
\end{aligned}
 \label{eq:ControlScheme},
\end{equation}

with $\rho$ setting the update timescale of $\lambda$ and the parameter $p$ controlling the rate of the control scheme (low $p$ value sets the first term in the parentheses close to $1$, so that $\lambda$ dynamics are slow).

\begin{figure}
\includegraphics[width=0.95\linewidth]{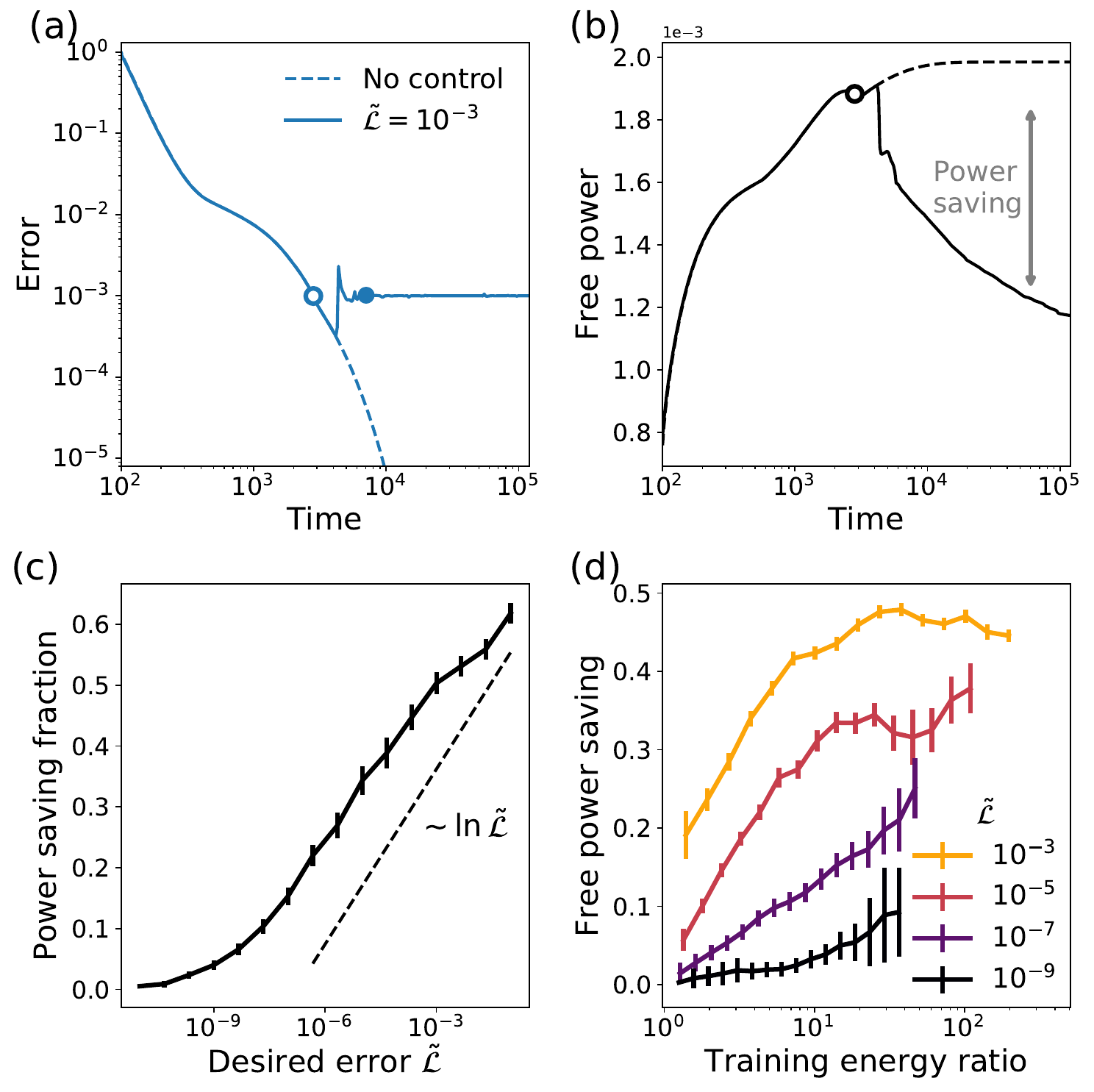}
\caption{Power-efficient solutions with dynamical control. a) Learning trajectories with our dynamical control scheme (full line) compared to simple learning without power minimization (broken line). We see that the controlled learning rapidly converges to a desired tolerable error level of $\mathcal{L}=10^{-3}$. b) The dynamically controlled system finds solutions that lower the free power dissipation compared to an early stopped training of the uncontrolled system at $\mathcal{L}=10^{-3}$ (open dot). The gray arrow signifies the saved power. c) The saved power given our control scheme compared to early stopping for different levels of tolerable error. We find that dynamical control can generate significantly more power-efficient solutions. d) However, to utilize the full benefit of power-efficient solutions, one needs to train the system for longer times, increasing the energy required to train the network.
\label{fig:Fig4}}
\end{figure}

To test this dynamical control scheme for learning with power optimization, we simulate training of $N=64$ nodes for regression tasks as before. We initialize the conductance values at their minimum $k_{\rm{min}}=10^{-3}$ and set $\alpha=0.03, \rho=1, p=0.02$. We find that the network trained with the $\lambda$ dynamical control scheme quickly converges on the desired error tolerance (Fig.~\ref{fig:Fig4}a, full line and closed circle). We compare these results with an ``early stopping algorithm," defined as follows. In this algorithm, we consider a learning network without power minimization ($\lambda=0$) (Fig.~\ref{fig:Fig4}a, dashed line). The network reaches  the desired error tolerance $\tilde{\mathcal{L}}=10^{-3}$ after some time (marked by the open circle on the dashed line in Fig.~\ref{fig:Fig4}a) that we call the ``early stopping time."  Note that our dynamical control scheme of of Eq.~\ref{eq:ControlScheme} reaches the same error at a time given by the solid circle on the solid line. Evidently the dynamical control scheme achieves lower power solutions compared with early stopping (Fig.~\ref{fig:Fig4}a). Once the dynamical control scheme reaches the time indicated by the solid circle in (Fig.~\ref{fig:Fig4}a), $\lambda$ stays constant but the system now trains itself at this value of $\lambda$, finally reaching a steady state at long times. As a result, the power advantage of this scheme (gray arrow in Fig.~\ref{fig:Fig4}b) improves over training time until it converges at some power value. 

We measure this power saving fraction at long training times and compare to the solution power for the early stopping algorithm for different error tolerances (Fig.~\ref{fig:Fig4}c). The power saving fraction is measured at $\tau=10^5$, in relation to the minimal power produced by the network given for the lowest possible conductance values $k_{\rm{min}}$. As higher error $\tilde{\mathcal{L}}$ is tolerated, the dynamical control scheme improves in comparison to the simple early stopping algorithm, saving an additional fraction of power that scales as $\ln \tilde{\mathcal{L}}$. We emphasize that this improvement in power is on top of utilizing the best conductance initialization. However, we note that gaining the full benefit of this power reduction requires longer training, possibly much longer than the early stopping time, meaning that the energy required to train the system is higher compared to the early stopping algorithm. This consideration means that in our dynamical control scheme there is a trade-off between training energy and the power-efficiency of the solution (Fig.~\ref{fig:Fig4}d). Such trade-off also depends on the error tolerance, but we find that if one is willing to spend $\sim5-10$ times the training energy compared to the early stopping algorithm, the network achieves most of the benefit of power reduction due to the dynamical control scheme. If the training energy is a major concern and constitutes a significant fraction of the energy expended by the network during its life, one should consider this trade-off for overall lowest power solutions. Finally, we note that our dynamical control scheme is not optimized. Choosing different parameters or another dynamical control scheme altogether may produce superior power saving at possibly lower training cost.

\section{Discussion}

In this work we studied how physical learning machines can be trained to adopt desired functions in power\textbackslash energy-efficient ways. We established that physical learning affects the power required to actuate the system given input signals. This power can be lowered by choosing better initialization schemes for the learning degrees of freedom, \textit{e.g.} initializing low conductances in electronic resistor networks.

We have also introduced a modified local learning rule that attempts to minimize both the error and power dissipation. We showed that this learning rule indeed lowers the power of obtained learning solutions in both simulations and experiments. This learning rule weights the importance of minimizing error vs. power dissipation, giving rise to a trade-off between the two. While improving power-efficiency at the expense of error (performance) may seem undesirable, very low error is typically not required and can even be infeasible in real learning situations. Therefore, one can often train learning networks to lower power solutions without much adverse effect (Appendix C). In our experiments, there is a natural noise floor and there is no point in striving for lower error than the floor. For these systems, power-efficient learning rules can improve solution power with little to no penalty in error. 

Finally, we have introduced a dynamical scheme for controlling the relative importance of error and power minimization to rapidly converge on power-efficient solutions with desired error tolerance. We find that such dynamical control can lead to lower power solutions. It is likely that an optimized version of such a dynamical control scheme could further reduce both the solution power and overall energy required to train the system. This is a subject of future study.

While we presented details of the analytical approach for the case of resistor networks, our theoretical arguments apply to other physical systems trained using coupled learning, such as mechanical spring networks (Appendix D). Neuromorphic computing often promises to improve power-efficiency by embedding learning algorithms in hardware, solving a major problem in modern power-hungry computational learning algorithms. While the hardware platform discussed here, self-learning electronic circuits, does indeed improve power efficiency, our work here focuses on how to achieve power-efficiency in the learning process itself. As a result, our power-efficient learning approach may be easily adaptable to other neuromorphic hardware systems that can perform self-learning, once they exist, to offer \emph{additional} power savings compared to only using efficient hardware.







\begin{acknowledgments}
We thank Purba Chatterjee, Marc Z. Miskin and Vijay Balasubramanian for insightful discussions and feedback. This research was supported by the U.S. Department of Energy, Office of Basic Energy Sciences, Division of Materials Sciences and Engineering award DE-SC0020963 (M.S.), the National Science Foundation via the UPenn MRSEC/DMR-1720530 and MRSEC/DMR-DMR-2309043 (S.D. and D.J.D.), and DMR-2005749 (A.J.L.), and the Simons Foundation (\# 327939 to A.J.L.). D.J.D. and A.J.L. thank CCB at the Flatiron Institute, as well as the Isaac Newton Institute for Mathematical Sciences under the program "New Statistical Physics in Living Matter" (EPSRC grant EP/R014601/1), for support and hospitality while a portion of this research was carried out. 

\end{acknowledgments}

\appendix

\section{Learning dynamics}

Here we provide more detail on the derivation of the learning dynamics, as well as how the free power required to actuate the network response changes during learning. Before tackling the question of the free power of a learning network, let us study the dynamics of the learning DOF, $k$, and the contrast, $\mathcal{C}$, due to the learning rule in Eq.~\ref{eq:LR1}. We assume there exists a solution of the learning degrees of freedom $k^*$ such that the contrast vanishes $\mathcal{C}(k^*)=0$ (this is the statement that the learning model is over-parameterized, so that the learning degrees of freedom can be trained to nullify the training error). Over-parameterization implies the existence of many connected solutions in $k$-space for which the contrast vanishes, and we denote by $k^*$ the solution obtained in practice by learning.  The contrast $\mathcal{C}$ is a complicated non-convex function of the learning DOF, but we can expand it around the solution $k^*$ to first non-vanishing order (second order):

$$
\mathcal{C}(k)\approx \frac{1}{2}(k-k^*)^T \mathcal{H} (k-k^*),
$$

where $\mathcal{H}\equiv\partial_k^2 \mathcal{C}(k^*)$ is the ``learning Hessian," \textit{i.e.} the Hessian of the contrast with respect to the learning DOF evaluated at the solution. Close enough to the learning solution $k^*$, we find that despite the explicit partial differentiation in Eq.~\ref{eq:LR1}, the learning dynamics are equivalent to gradient descent on the contrast~\cite{stern2021supervised}. Therefore, if we absorb the learning rate into the definition of the time unit, the weight dynamics are given by $\dot{k}=-\nabla_k\mathcal{C}=-\mathcal{H}(k-k^*)$. This leads to simple exponential decaying dynamics. If we set the initial condition at $k(t=0)\equiv k^0$, then

\begin{equation}
\begin{aligned}
k(t)=k^*+e^{-\mathcal{H}t}(k^0-k^*)
\end{aligned}
 \label{eq:WDyn}.
\end{equation}

Setting the time propagator operator $U(t)\equiv e^{-\mathcal{H}t}=U^T$, we can use this result to obtain the decaying dynamics of the contrast:

\begin{equation}
\begin{aligned}
\mathcal{C}(t)=\frac{1}{2}(k^0-k^*)^T U\mathcal{H}U(k^0-k^*)
\end{aligned}
 \label{eq:C0}.
\end{equation}

While these results are consistent with simple exponential decay of the learning DOF $k$ and contrast $\mathcal{C}$, one complication typically arises for over-parameterized learning. We have seen before that the learning Hessian in over-parameterized learning machines tends to be low-rank (with the number of non-zero eigenvalues equal to the number of training tasks)~\cite{stern2023physical}. As the learning Hessian has zero eigenvalues it is not invertible. In the eigen-directions of these vanishing eigenvalues there are no dynamics, as can be explicitly seen by rotating the frame into the coordinate system that diagonalizes $\mathcal{H}$. The learning dynamics are agnostic to components of $k$ in the the large null-space of $\mathcal{H}$. We can plug these results in Eq.~\ref{eq:EF0} to obtain the free state power dynamics 

\begin{equation}
\begin{aligned}
\dot{P}^F(k)=(k^*-k^0)^TU\mathcal{H} \partial_k P^F(k^*)
\end{aligned}
 \label{eq:EF01}.
\end{equation}

As only $U(t)$ depends on time, this ODE can be integrated to find that the free state power exponentially saturates to a value 

\begin{equation}
\begin{aligned}
P^F(t\rightarrow\infty)= P^F(t=0) + (k^*-k^0)^T A^T A \partial_k P^F(k^*)
\end{aligned}
 \label{eq:EF02},
\end{equation}

where $A$ is a projection matrix, projecting weight vectors into the stiff (\textit{i.e.} non-null) subspace of $\mathcal{H}$. Here we see again that the power can increase or decrease during learning, depending on the alignment between the gradient of the free state power and the direction of weight dynamics.

\begin{figure}
\includegraphics[width=0.95\linewidth]{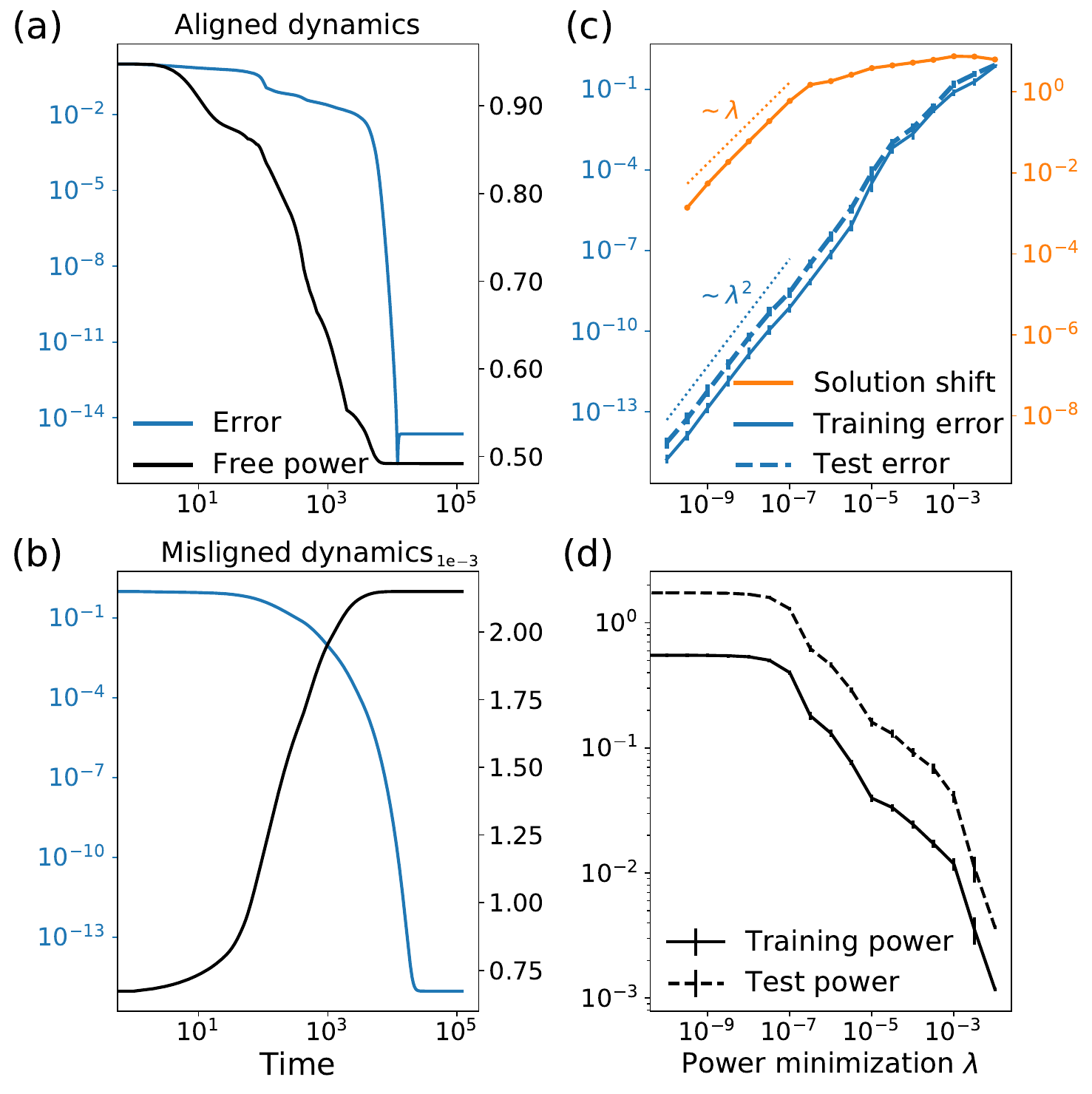}
\caption{Learning dynamics with power minimization. a) When the gradients of the error and free power align, both are reduced by learning, and the error undershoots its steady state value before relaxing back to it. b) In contrast, when the error and free power gradients do not align, learning increases the free power while smoothly reducing the error to its final value. c) as the power minimization amplitude $\lambda$ is increased, the learned solution $k^*_\lambda$ linearly displaces from the limiting solution $k^*_{0^+}$. The error increases quadratically with $\lambda$ both for the training set and also for the test set. d) The free power of the learned solution is decreased by learning at higher $\lambda$ for both the training and test set inputs.
\label{fig:FigSI1}}
\end{figure}

We now discuss the modified learning dynamics that minimizes both error and power (Eq.~\ref{eq:CFLD}). In the main text we showed that these learning dynamics lead to exponentially decaying weight solutions (Eq.~\ref{eq:WExpSol}) and associated error and power dynamics given by Eq.~\ref{eq:Solutions}. The dynamical trajectories given these error\textbackslash power dynamics follow two different prototypes, depending on the sign of $\phi \equiv -\partial_k\mathcal{C}^T \partial_k P^F$. For $\phi<0$ (where the contrast gradient is aligned with the power gradient), the contrast undershoots the infinite time limit, getting arbitrarily close to $\mathcal{C}=0$ before rebounding exponentially to $\mathcal{C}(t\rightarrow\infty)$ (Fig.~\ref{fig:FigSI1}a). This scenario is common when initializing the network with high conductance values. For $\phi>0$ (\textit{i.e.} anti-alignment of the contrast and power gradients), dynamics tend to increase the power, and we see analytically regular dynamics, where the contrast smoothly decays exponentially to its terminal value $\mathcal{C}(t\rightarrow\infty)$ (Fig.~\ref{fig:FigSI1}b). This scenario is common in flow\textbackslash resistor networks initialized at low conductance values. In Fig.~\ref{fig:FigSI1}c, we verify the argument laid out in the main text that the solution $k^*_\lambda-k^*_{0^+} \sim \lambda$. We also show as before that the error grows quadratically with $\lambda$. Crucially, the arguments for the error are relevant not only for the training set (regression examples used to train the network) but also for test examples the network had not seen previously, whose error also scales quadratically in $\lambda$. More information about the regression tasks, as well as the training and test sets, is available in Appendix B. Similarly to the error, the arguments about the power of the obtained solution are valid for both the training and test sets, so that our modified learning dynamics reduces both of them (Fig.~\ref{fig:FigSI1}d).


\section{Physical learning tasks}

Here we describe the regression tasks explored numerically in the main text. We simulated linear resistor networks with $N=64$ whose structure is derived from jammed 2-dimensional packings~\cite{goodrich2015principle}. We randomly choose $2$ edges as input edges and another $2$ as output edges (see Fig.~\ref{fig:Fig1}a). The input and output voltage drops are noted by the vectors $\Delta V_i, \Delta V_o$, respectively. The network is trained to perform regression recovering a linear relation

\begin{equation}
\begin{aligned}
\Delta V_o + \epsilon = \sum_{i} \tilde{A}_{oi} \Delta V_i
\end{aligned}
 \label{eq:regress}.
\end{equation}

Here, the $2\times 2$ matrix $\tilde{A}_{oi}$ contains the desired function parameters and $\epsilon$ a possible addition of white noise. Since we train a linear resistor network, the functional relation between the input and output voltage drops is always linear $\Delta V_o = \sum_{i} A_{oi} \Delta V_i$, and the correct matrix relation $\tilde{A}_{oi}$ is supposed to be recovered by learning. The values for the desired matrix were randomly chosen from the distribution

\begin{equation}
\begin{aligned}
\tilde{A} \sim \begin{pmatrix} 0.2 & 0.3  \\  0.1 & 0.5 \end{pmatrix}&\ + \ 0.1 \mathcal{N}(0, 1)^{2\times 2}
\label{eqn:AMat}
\end{aligned}
\end{equation}

We trained these networks in many realizations of geometry, choice of input\textbackslash output edges and $\tilde{A}_{oi}$. To train each realization of the problem we sampled $20$ training examples $\Delta V_i^{\rm{Training}} \sim U(0,1)^2$ and corresponding outputs $\Delta V_o^{\rm{Training}} = \sum_{i} \tilde{A}_{oi} \Delta V_i^{\rm{Training}} + \epsilon$. Note that the scale of the input voltage drops determines the scale of power dissipation in the free state $P^F\sim \Delta \bar{V_i}^2$. In the main text we looked at noiseless regression problems with $\epsilon=0$, for which the network can find exact solutions with zero error. In Appendix C we study a case with finite label noise $\epsilon=10^{-3}$. The training examples are sampled randomly during training and used to define the free and clamped states in the iterative learning process. Apart from the $20$ training examples, we also sampled $100$ test examples from a wider distribution $V_i^{\rm{Test}}\sim \mathcal{N}(0,1)$ and their associated desired outputs. The test points are not used during the learning process but help in verifying that the network can generalize. In our work, the test set is interesting also for showing the power-efficient property of the solutions generalizes beyond the training set (Fig.~\ref{fig:FigSI1}d).

\section{Power minimization for limited accuracy tasks}

The numerical results in the main text were limited to tasks that can, in principle, be learned perfectly by the learning machine. In such cases there exist solutions with no error $\mathcal{L}(k^*)=0$, as discussed in Appendix A. There are, however, cases in which it is impossible to obtain solutions with zero error. The typical example is when the training set does not capture all of the information contained in the broader data (or the test set). There are also cases where it is impossible to find solutions that nullify the error even on the training set. This can occur due to under-parameterization (too few learning degrees of freedom to learn the task), an insufficiently expressive model (\textit{e.g.} a linear network cannot represent non-linear relations) and noise in the learning process~\cite{mehta2019high}.

We will first consider the case where the system ends up in a local minimum with $\mathcal{L}>0$. From the definition of coupled learning, we know that if the loss is finite $\mathcal{L}>0$, so is the contrast $\mathcal{C} > 0$. A minimum of the coupled learning dynamics $k^\dagger$ in such a case would have finite error and contrast values $\mathcal{L}(k^\dagger), \mathcal{C}(k^\dagger)$. Nonetheless, we can still perform a quadratic approximation around the contrast minimum $k^\dagger$ similar to Appendix A, where the constant term $\mathcal{C}(k^\dagger)$ is retained: 
$$
\mathcal{C}(k)\approx \mathcal{C}(k^\dagger) + \frac{1}{2}(k-k^\dagger)^T \mathcal{H} (k-k^\dagger).
$$

Using this expansion, we can redo the derivation of Section III to find the steady state solution error and power when a finite power minimization amplitude $\lambda$ is applied in Eq.~\ref{eq:CFLD}:

\begin{equation}
\begin{aligned}
\mathcal{C}(\lambda)&\approx \mathcal{C}(k^\dagger) + \frac{1}{2}\lambda^2 s^T\mathcal{H}s \\
P^F(\lambda)& \approx  P^F_{0^+}+ \lambda (\partial_k P_{0^+}^{F})^T (\mathcal{H}+\lambda H)^{-1} \partial_k P_{0^+}^{F}
\end{aligned}
 \label{eq:SolutionsInfSI}
\end{equation}

Comparing these expressions to Eq.~\ref{eq:SolutionsInf}, we see that the power behavior stays the same. We also see that the error shift is the same, scaling as $\lambda^2$, but now there is a finite contrast floor $\mathcal{C}(k^\dagger)$ associated with a finite error. The trade-off between error and power is still maintained although in this case it may be much more favorable. For small enough $\lambda$, $\frac{1}{2}\lambda^2 s^T\mathcal{H}s \ll \mathcal{C}(k^\dagger)$ and so the contrast (and error) is nearly unaffected by the power minimization. As a result, we can apply a finite power minimization parameter $\lambda$, reducing the solution power at virtually no penalty. Power minimization is thus particularly useful for problems in which zero error solutions are not possible. 

\begin{figure}
\includegraphics[width=0.95\linewidth]{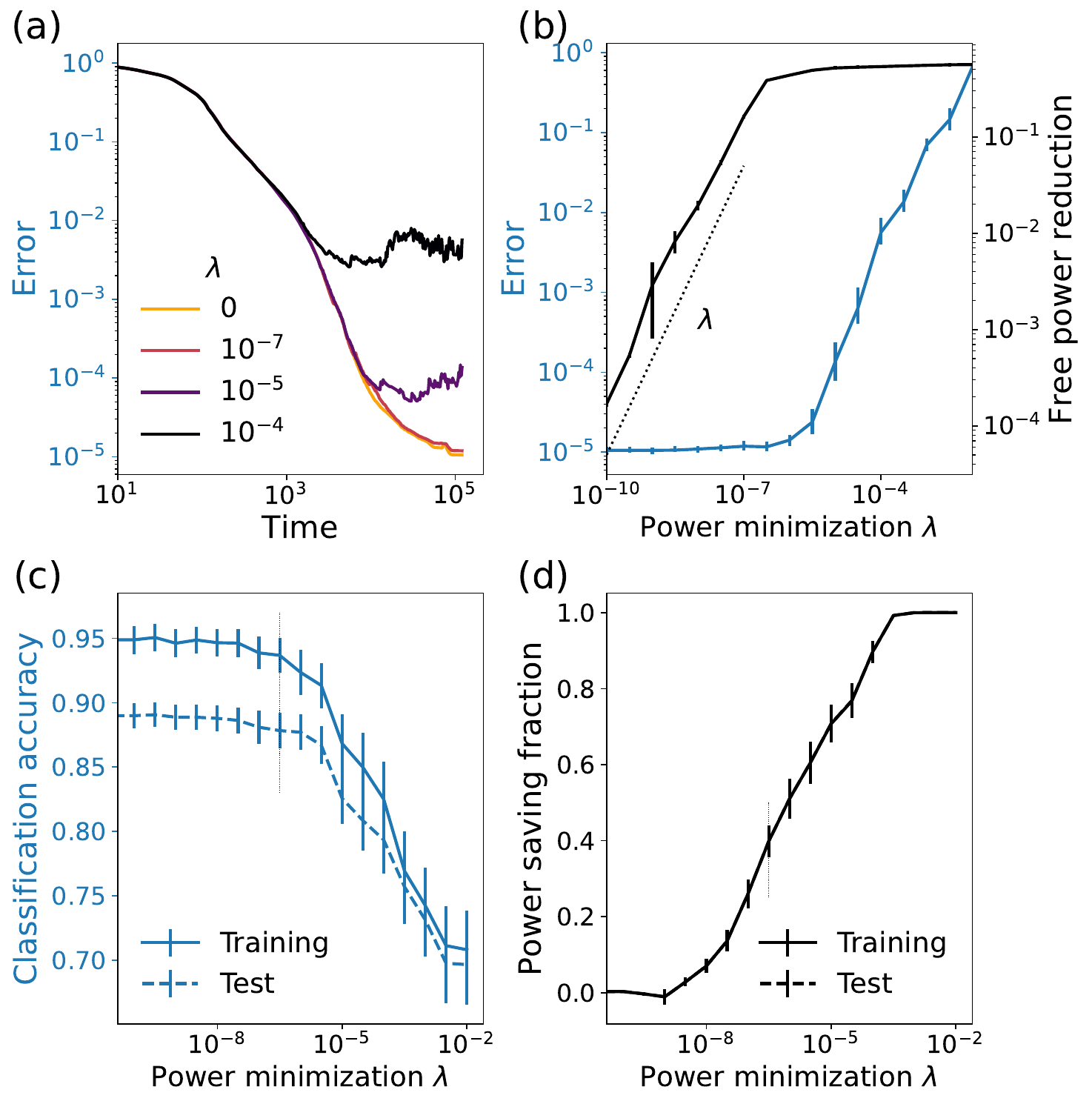}
\caption{Power reduction with little error\textbackslash accuracy loss in regression and classification problems. a) Error trajectories for regression task with label noise, so that the minimum possible error is $\mathcal{L}\approx10^{-5}$. As long as the power minimization parameter is small enough $\lambda<10^{-7}$, the error is largely unaffected. b) However, the free power is still reduced by increasing $\lambda$ even in the range where the error is unaffected. c) Similar results are found for a classification task on the Iris dataset, as increasing the power minimization amplitude $\lambda$ decreases accuracy (\textit{i.e.} increases error), but only beyond a finite value of $\lambda$. d) Increasing $\lambda$ in such classification problems decreases the free power for both the training and test sets.
\label{fig:FigSI2}}
\end{figure}

To verify these considerations, we simulated physical learning in $N=64$ networks on regression and classification tasks (Fig.~\ref{fig:FigSI2}). Excess noise was added to the regression labels (outputs) in the training and test sets, sampled from a distribution $\Delta V_o \sim \sum_{i} \tilde{A}_{oi} \Delta V_i + \epsilon$, with $\epsilon=10^{-3}$ (see Appendix B). The simulated networks can successfully learn these tasks, reducing the error to some finite value $\mathcal{L}\approx10^{-5}$ (Fig.~\ref{fig:FigSI2}a). When a adding small power minimization $\lambda<10^{-7}$, the learning trajectories are almost unchanged and the error is nearly unaffected. When larger power optimization is introduced, the error starts increasing beyond the error floor (Fig.~\ref{fig:FigSI2}b). At the same time, we observe that the free power is decreased linearly at finite $\lambda$ as seen before. These results show that in noisy cases, such as seen in physical learning experiments, power reduction can be achieved at no expense in error up to a certain point.

Another case where this result is particularly relevant is in classification problems, where we would like to assign discrete labels to inputs. 
A standard example for such tasks is the classification of Iris specimens based on measurements of lengths of their petals and sepals~\cite{Iris}. We have previously shown our flow\textbackslash resistor networks can successfully learn to classify the Iris dataset, as well as could be expected from linear network models, in simulations~\cite{stern2023physical} and experiments~\cite{dillavou2022demonstration}. In discrete classification tasks, we are typically not concerned with the mean squared error, but with a measure of accuracy given by a discrete choice of the label based on the network response; excellent classification is possible even at relatively high values of the mean squared error. Therefore, it may be possible to induce power optimization without penalty in classification accuracy. To test this idea, we simulated training of our $N=64$ node networks to classify the iris dataset (a detailed description of the training protocol can be found in~\cite{dillavou2022demonstration}). Training at different power minimization amplitudes $\lambda$ in the range $10^{-10}<\lambda<10^{-2}$, we find that the classification accuracy (for the training and test sets) is not affected by power minimization until $\lambda\approx 10^{-7}$ (Fig.~\ref{fig:FigSI2}c). At the same time, the solution free power is significantly reduced starting at $\lambda > 10^{-8}$, showing power saving (in this case, by a factor $\sim 2$) is possible at little penalty in accuracy (Fig.~\ref{fig:FigSI2})d). 

We turn now to another case in which tasks cannot be learned perfectly, this time due to the existence of noise. In any real physical learning machine noise in measurement and learning DOF updates will lead to a nonzero error floor associated with physical learning. This is true even for tasks that in the absence of noise could be learned with no error. In such setting, the random noise pushing the system away from the zero contrast (and error) minima implies physical learning behaves as a high dimensional Ornstein-Uhlenbeck process~\cite{vatiwutipong2019alternative} in the space of the learning DOF. The instantaneous values of the learning DOF are then sampled from a normal distribution centered around $k_\lambda^*$ of Eq.~\ref{eq:other}, with a standard deviation scaling with the white noise amplitude $\sigma$~\cite{gardiner1985handbook}:

\begin{equation}
\begin{aligned}
k_{\lambda,i}(\sigma) &\sim k_{\lambda,i}^*(\sigma=0) + \frac{\sigma}{\sqrt{2\theta_{\lambda,i}}} \mathcal{N}(0,1)
\end{aligned}
 \label{eq:KSSI}
\end{equation}

where $\theta_{\lambda,i}$ are the eigenvalues of the matrix $\mathcal{H}+\lambda H$. In other words, the noise induces the conductances to explore a vicinity of the solution $k_\lambda^*$, whose size depends on the noise amplitude $\sigma$ and the curvature given by the eigenvalues $\theta_{\lambda,i}$. We can take this distribution of values of the learning DOF and plug it in the equation for the contrast (Eq.~\ref{eq:RegCont}), finding the distributions of this quantity. 

\begin{equation}
\begin{aligned}
\mathcal{C}_{\lambda}(\sigma) &\sim \frac{1}{2}\lambda^2 s^T\mathcal{H}s + \sum_i\frac {\lambda \sigma s_i}{\sqrt{2 \theta_{\lambda,i}}}\mathcal{N}_i(0,1) + \\
&+ \sum_i\frac {\sigma^2}{4 \theta_{\lambda,i}}\mathcal{N}_i^2(0,1)
\end{aligned}
 \label{eq:KSSI2}
\end{equation}

This can similarly be done for the free power saving for given $\lambda$. The average contrast induced by the noise, as well as the average free power saving, can be deduced by taking the expectation value over these distributions. Here, note that the the expectation values of these normal distributions are $\langle\mathcal{N}_i(0,1)\rangle=0, \langle\mathcal{N}_i^2(0,1)\rangle = 1$, so that we are left with

\begin{equation}
\begin{aligned}
&\langle \mathcal{C}_{\lambda}(\sigma) \rangle \approx \frac{1}{2}\lambda^2 s^T\mathcal{H}s + \sum_i\frac {\sigma^2}{4 \theta_{\lambda,i}} \\
& \langle P^F_\lambda- P^F_{0^+} \rangle(\sigma) \approx  \lambda (\partial_k P_{0^+}^{F})^T (\mathcal{H}+\lambda H)^{-1} \partial_k P_{0^+}^{F}
\end{aligned}
 \label{eq:KSSI3}.
\end{equation}

We find that if we know the noise scale $\sigma$, measuring the average contrast value allows $\langle \mathcal{C}_{0}(\sigma) \rangle$ allow us to glean information about the effective average curvature of the contrast near the learning solution. Note that the learning DOF diffuse freely in the space of zero contrast solutions, so the effective curvature is associated with the typical slopes of the contrast leaving the zero manifold. Overall, we see that the free power saving is on average the same as in the case with no noise (up to second order terms in $\lambda$). However, the contrast now has a finite added term due to the exploration of values of the learning DOF beyond the minimum $k^*_\lambda$. This means additive white noise has a similar effect to the finite contrast floor discussed earlier; finite power optimization $\lambda$ can reduce the free power while having nearly no affect on the contrast (or error) up to a certain scale.

\section{Power minimization in mechanical spring networks}

In this work, we presented general arguments on how local learning rules could balance minimizing the error and power of obtained physical learning solutions, giving rise to a trade-off between the two. However, in the main text we only tested these ideas numerically and experimentally in resistor networks. Here, we show in simulations that these arguments apply similarly to physical learning systems governed by different physics, \textit{i.e.} an elastic network of harmonic springs (Fig.~\ref{fig:FigSI3}a). 

Elastic networks have been studied as a nonlinear substrate for physical learning~\cite{pashine2019directed,hexner2019effect,hexner2019periodic,stern2020continual,stern2020supervised,hexner2021adaptable,arinze2023learning}. Specifically, coupled learning can train spring networks networks to perform desired tasks by modifying the spring constants or rest lengths~\cite{stern2021supervised}. The physical cost function naturally minimized by such networks is the elastic energy $E$:

\begin{equation}
\begin{aligned}
E = \frac{1}{2}\sum_i k_i (r_i - \ell_i)^2
\end{aligned}
 \label{eq:SpE},
\end{equation}

where $k_i$ is the spring constant of spring $i$, $\ell_i$ its rest length, $r_i$ the Euclidean distance between the nodes connected by the spring, and the energy is summed over all individual springs. For a spring network with adaptive spring constants, the local learning rule is:

\begin{equation}
\begin{aligned}
\dot{k}_i &=-\alpha\eta^{-1}\frac{\partial}{\partial k_i} [E^C-E^F] =\\
& = -\frac{1}{2}\alpha \eta^{-1} [(r_i^C - \ell_i)^2 - (r_i^F - \ell_i)^2],
\end{aligned}
 \label{eq:spl}
\end{equation}

where $r_i^F, r_i^C$ are the distances between nodes separated by spring $i$ in the free and clamped state, respectively. More details on the derivation of this learning rule can be found in ref.~\cite{stern2021supervised}. To see if spring networks can trained to adopt low energy solutions, \textit{i.e.} spring configurations for which the desired state is easy (takes little energy) to actuate, we add a local energy minimization term with amplitude $\lambda$, similarly to Eq.~\ref{eq:CFLD}:

\begin{figure}
\includegraphics[width=0.95\linewidth]{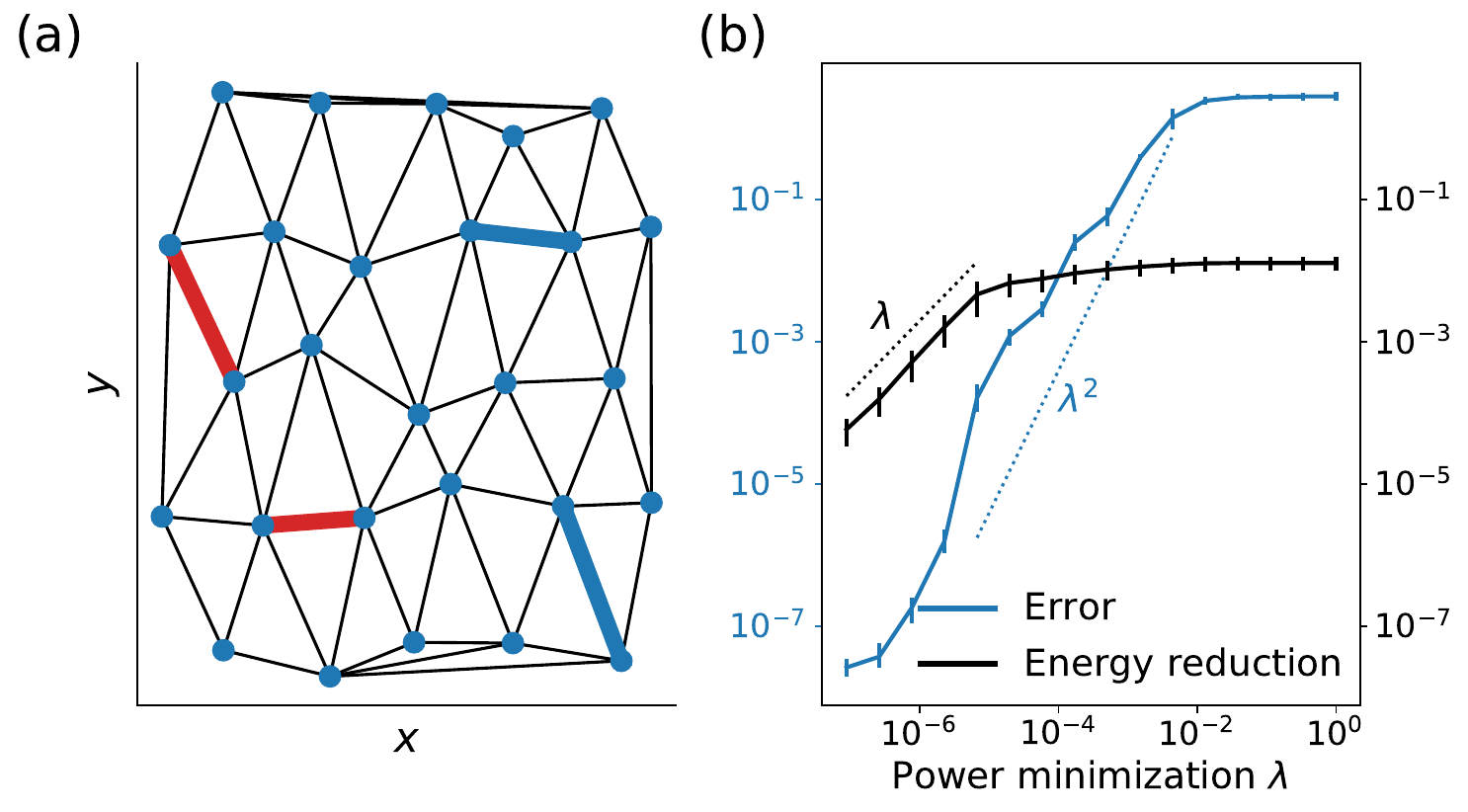}
\caption{Energy-efficient learning in mechanical spring networks. a) A mechanical spring network, each edge corresponding to a spring with adaptive stiffness $k$. Such networks are trained for allostery tasks, so that prescribed strains at input edges (red) lead to desired strains at output edges (blue). b) As seen for flow networks, including a power minimization term in the local learning rule leads to a trade-off between error and power, also having the same scaling behavior.
\label{fig:FigSI3}}
\end{figure}

\begin{equation}
\begin{aligned}
\dot{k}_i &=-\alpha\eta^{-1}\frac{\partial}{\partial k_i} [E^C-(1-\lambda)E^F]
\end{aligned}
 \label{eq:splm}
\end{equation}

We simulate this modified learning algorithm on unstrained spring network with $N=27$ nodes as shown in Fig.~\ref{fig:FigSI3}a. These networks are trained for allostery tasks, in which we apply prescribed relative strains $0.2$ (randomly choosing contraction or extension) and desire particular strain values at another two random bonds ($0.05$ or $0.03$, randomly choosing contraction or extension). With no energy minimization, $\lambda=0$, coupled learning generally succeeds in training these networks to a numerical normalized error floor of $\mathcal{L}\sim 10^{-8}-10^{-7}$. As we increase the power minimization amplitude $\lambda$, we observe that the error increases as $\lambda^2$ and the solution energy reduced as $\lambda$ (Fig.~\ref{fig:FigSI3}b), as predicted by Eq.~\ref{eq:SolutionsInf} and observed in simulations of resistor networks. These results show our approach to physical learning of power efficient solutions can be employed beyond linear resistor networks.




\end{document}